\documentclass[journal=ancac3,manuscript=article]{achemso}

\usepackage{chemformula}
\usepackage[T1]{fontenc} 
\usepackage{hyphenat} 
\usepackage{amsmath,amssymb}
\usepackage{bbding}
\usepackage[notextcomp]{stix}
\usepackage{placeins}

\author{Jacopo Vialetto}
\affiliation{Department of Chemistry, University of Florence, via della Lastruccia 3, Sesto Fiorentino, I-50019 Firenze, Italy}
\alsoaffiliation{Laboratory for Soft Materials and Interfaces, Department of Materials, ETH Z{\"u}rich, Vladimir-Prelog-Weg 5, 8093 Z{\"u}rich, Switzerland}
\email{jacopo.vialetto@unifi.it}

\author{Shivaprakash N. Ramakrishna}
\affiliation{Laboratory for Soft Materials and Interfaces, Department of Materials, ETH Z{\"u}rich, Vladimir-Prelog-Weg 5, 8093 Z{\"u}rich, Switzerland}

\author{Lucio Isa}
\affiliation{Laboratory for Soft Materials and Interfaces, Department of Materials, ETH Z{\"u}rich, Vladimir-Prelog-Weg 5, 8093 Z{\"u}rich, Switzerland}

\author{Marco Laurati}
\affiliation{Department of Chemistry, University of Florence, via della Lastruccia 3, Sesto Fiorentino, I-50019 Firenze, Italy}
\email{marco.laurati@unifi.it}

\title{Effect of particle stiffness and surface properties on the non-linear viscoelasticity of dense microgel suspensions}


\begin{document}

\begin{abstract}
Particle surface chemistry and internal softness are two fundamental parameters in governing the mechanical properties of dense colloidal suspensions, dictating structure and flow, therefore of interest from materials fabrication to processing. Here, we modulate softness by tuning the crosslinker content of poly(N-isopropylacrylamide) microgels, and we adjust their surface properties by co-polymerization with polyethylene glycol (PEG) chains, controlling adhesion, friction and fuzziness.
We investigate the distinct effects of these parameters on the entire mechanical response from restructuring to complete fluidization of jammed samples at varying packing fractions under large-amplitude oscillatory shear experiments, and we complement rheological data with colloidal-probe atomic force microscopy to unravel variations in the particles' surface properties.
We find that surface properties play a fundamental role at smaller packings; decreasing adhesion and friction at contact causes the samples to yield and fluidify in a lower deformation range. Instead, increasing softness or fuzziness has a similar effect at ultra-high densities, making suspensions able to better adapt to the applied shear and reach complete fluidization over a larger deformation range.
These findings shed new light on the single-particle parameters governing the mechanical response of dense suspensions subjected to deformation, offering synthetic approaches to design materials with tailored mechanical properties.
\end{abstract}

\bigskip

\section{Introduction}
Dense suspensions of colloidal particles are at the core of a variety of materials and processes, being fundamental components in formulations for surface coatings, fluids for oil extraction, additive manufacturing, cosmetics, food-grade materials, etc.. \cite{Scheffold2020,Ness2022}
In addition to the particle surface properties, which dictate the strength of the main interparticle interactions at play (e.g., electrostatic, steric, van der Waals) the internal degree of softness of the colloids dramatically affects their behavior in crowded states, and it is used as a precisely controllable parameter to tune their structural and rheological properties. \cite{Vlassopoulos2014,Scotti2022} 
This is of great interest for the fundamental understanding of the rich phase behavior of a broad range of soft systems, including biological ones, \cite{Li2017,Hernando-Perez2019} and for the design of functional materials with tuneable mechanical responses, as required in applications ranging from optics to viscosity modifiers and biocompatible carriers. \cite{Hu2003,Bhattacharjee2015,Douglas2017,Miksch2022,Jung2024}

Among soft colloids, microgel particles made of swellable polymer networks offer multiple opportunities to achieve the desired functionalities in dense systems. When dispersed in a good solvent they are highly deformable, compressible, and can interpenetrate with neighbouring particles, \cite{Conley2017,Conley2019} allowing one to easily reach states with effective volume fractions that exceeds the hard-sphere limit, and can be higher than 1. \cite{VanderVaart2013,Vlassopoulos2014,Zhou2023}.
The extent of deformation can be tuned at the synthesis level by varying the crosslinker content, or by using polymers responsive to external inputs (e.g., temperature, pH, ionic strength). \cite{Scotti2022} This has a profound effect on the phase diagrams, flow and rheological properties of microgels in densely packed states, both in suspensions \cite{Carrier2009,Paloli2013,Pellet2016,Bergman2018,Conley2019,Scotti2020PRE,nikolov2020} and in two-dimensions (i.e., when microgels are compressed on a fluid interface). \cite{Scotti2019,Vialetto2021,Vialetto2022JCIS} Tuning their internal elasticity allows, for example, to shift the transition between liquid and crystalline state to higher effective concentrations, \cite{Scotti2019,Scotti2020PRE}, to crystallize suspensions even in the presence of very high polydispersity, \cite{Iyer2009,Scotti2017} or to modulate the material viscoelastic properties such as storage modulus and yield strain. \cite{Scotti2020,Saisavadas2023}

The rheological response of microgel suspensions can be further modified acting on the interparticle contacts, consequently tuning the resulting interactions.
For example, temperature variations in the case of poly(N-isopropylacrylamide) (pNIPAM) microgels allow inducing transitions from a more hydrophilic to a more hydrophobic surface, triggering the emergence of attractive interactions. \cite{Romeo2010,Chaudhary2021} Alternatively, different polymers on the surface can be exploited to modulate the extent of adhesion, friction and interdigitation between particles. \cite{Menut2012,Urayama2014,Moreno-Guerra2019,Lara-Pena2021,Guerron2021,Ruiz-Franco2023} However, surface effects, albeit of utmost importance to fully comprehend the behavior of soft-particle systems, are rarely linked to the rheological properties of dense microgel suspensions.

In this work we focus on the linear and non-linear rheological behavior of dense pNIPAM microgel suspensions in the jammed state under oscillatory shear, with specific emphasis on the role of the internal degree of softness and the influence of particle surface properties. Although their viscoelastic response, with particular emphasis on the yielding transition, have been already investigated previously, \cite{Carrier2009,Koumakis2012,VanderVaart2013,Pellet2016,Saisavadas2023}
the full dynamics of yielding and the path towards complete fluidization under large amplitude oscillatory shear (LAOS) are still largely unexplored. Furthermore, the effects of anharmonic contributions on the entire fluidization process have not been fully considered. In particular, we herein analyze the effect of crosslinking density and that of adding linear polyethylene glycol (PEG) chains on the particle surface to tune their mechanical response by affecting preferentially the internal particle elasticity\cite{nikolov2020}, or the interactions at contact.
We take advantage of the sequence of physical processes (SPP) \cite{Rogers2017,Donley2019} approach to analyse LAOS measurements and gain insights on the viscoelastic non-linear response during progressive fluidization of the samples at increasing applied deformations. By looking at LAOS data we, in particular, disclose the influence that particle softness has in governing their rheological behavior in very high density states. Additionally, to corroborate the rheological data, we employ colloidal probe atomic force microscopy (CP-AFM) to quantify the decrease in the particles' adhesive and frictional forces mediated by the incorporation of PEG comonomers on the microgel surface. 

\section{Results and discussion}

In Figure \ref{fig1}a we report a schematic of the microgels synthesized and analyzed in this work. We investigated pNIPAM microgels with 5 and 1 mol \% crosslinker (identified as pN5 and pN1, respectively), as well as pNIPAM-PEG microgels with 5 and 1 mol \% crosslinker (pN-PEG5 and pN-PEG1, respectively). Details on the synthesis protocols are in the Materials and Methods section. Importantly, for the synthesis of our pNIPAM-PEG microgels we used a semi-batch protocol in order to increase the relative content of PEG chains in the outer shell, and thus on the surface, of the particles.
We first initiated the radical precipitation polymerization of NIPAM monomers in the reaction flask to produce polymeric cores containing only pNIPAM. Only afterwards, we injected PEG monomers (polyethylene glycol methyl ether methacrylate, PEGMA, m.w.: 13000) at a fixed rate in the presence of the pNIPAM nuclei. The use of a semi-batch protocol, together with the lower reaction constant of long PEGMA chains with respect to NIPAM, is believed to effectively accumulate the linear PEG comonomers on the microgels' surface. \cite{Motlaq2018,EsSayed2019}

Dynamic light scattering is used to quantify the hydrodynamic diameter (d$_h$) of the different microgels as a function of temperature (Figure S1). Both the sharpness of the transition and the swelling ratio (see also Table \ref{tab1}) are not affected by the addition of PEG, and the swelling ratio only depends on the crosslinker content. This is in accordance with a particle architecture composed of PEG chains located mainly on the particle periphery that do not influence the thermal response of the pNIPAM network. \cite{EsSayed2019,Agnihotri2020} The fact that the deswelling is nearly the same for particles with the same crosslinker content (Figure S1b) supports the assumption that PEG chains do not modify the effective crosslinking density and therefore the single-particle internal elasticity. 

\subsection{Non-linear rheology of pNIPAM microgels}

\begin{figure}[t!]
\centering
\includegraphics[scale=1]{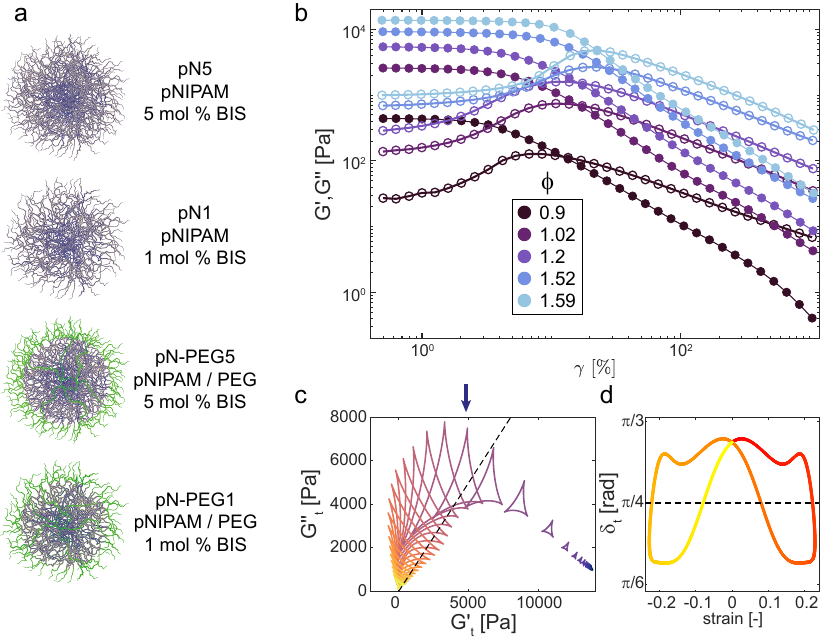}
\caption{\small \textbf{Rheological behavior of pNIPAM and pNIPAM-PEG microgels.} a) Sketch of the microgels synthesized and investigated in this work. b) Storage ($G'$) and loss ($G''$) moduli from strain amplitude sweeps at $\omega = 10$ rad/s of pN5 microgels at increasing effective volume fraction ($\phi$). c) Instantaneous values of the time-resolved viscoelastic SPP moduli $G'_t$ and $G''_t$ for pN5 microgels at $\phi = 1.59$. Color-coding from blue to yellow indicates increasing $\gamma \%$ in the amplitude sweep experiment. The dashed line marks the yielding threshold $\delta_t = \pi/4$. d) Plot of the instantaneous phase angle $\delta_t$ corresponding to the data with $\gamma = 22.7\%$ in (c) indicated by the arrow. Color-coding from red to yellow shows the time within one oscillation cycle.}
\label{fig1}
\end{figure}

We begin by analyzing the rheological response of pN5 microgels at varying effective volume fractions ($\phi$) beyond close packing, i.e. starting at 0.9. The linear viscoelastic moduli obtained from frequency sweeps (Figure S2a) indicate a solid-like response for all samples, with the storage modulus $G'$ always higher than the loss modulus $G''$. The plateau viscoelastic moduli increase linearly with $\phi$ (Figure S2b), as expected for pure pNIPAM microgels at comparable concentrations, \cite{Scheffold2010,VanderVaart2013,Lara-Pena2021} and  as ascribed to the finite softness of their cores. \cite{Menut2012,romeo2013,Pellet2016}
We note that all samples are well within the jamming limit as estimated by Pellet and Cloitre. \cite{Pellet2016} The microgels are therefore expected to be highly deformed, compressed and possibly interpenetrated. \cite{Conley2019} 

We then investigate the non-linear response in LAOS experiments. In Figure \ref{fig1}b we report $G'$ and $G''$ measured in first harmonic approximation, for an oscillation frequency $\omega = 10$~rad/s. For all samples, the moduli are approximately constant up to strain amplitudes ($\gamma$) of a few percent, while at larger applied strains the material response changes from predominantly solid-like ($G' > G''$) to liquid-like ($G' < G''$), exhibiting a weak $G''$ overshoot behavior that has been recently associated to a continuous transition from recoverable to unrecoverable deformation. \cite{Donley2020} This response agrees with results typically reported for microgels, \cite{Carrier2009,VanderVaart2013,Pellet2016,Saisavadas2023} hard-core soft-shell particles, \cite{Brader2010,Koumakis2012} and other colloidal systems with varying softness. \cite{Vlassopoulos2014}

We used the time-resolved SPP analysis\cite{Rogers2017,Donley2019} (see Materials and Methods) to get detailed insights into the non-linear behavior of the samples as a function of the applied strain and to obtain information on the dynamics of the mechanical response during the yielding transition.
Figure \ref{fig1}c shows Cole-Cole plots of the variation of the instantaneous moduli $G'_t$ and $G''_t$ during the cycles for increasing $\gamma$ (color-coded from blue to yellow) for pN5 microgels at $\phi = 1.59$. At low applied strains $G'_t >> G''_t$ for the entire cycle, as expected for a solid-like response. Upon increasing $\gamma$, $G'_t$ decreases, $G''_t$ increases, and both moduli change significantly during the oscillation cycle. For this sample, we observe that $G''_t$ becomes higher than $G'_t$, corresponding to a partial yielding of the material within the cycle,\cite{Donley2019} starting from $\gamma = 22.7\%$ (arrow in Figure \ref{fig1}c). 
The instantaneous moduli are then used to calculate the instantaneous phase angle $\delta_t$ (eq. \ref{delta_t}) as defined by Donley et al.,\cite{Donley2019} which is used to identify the point within the cycle when the material response changes from primarily elastic, $\delta_t < \pi ⁄4$, to primarily viscous, $\delta_t > \pi ⁄4$, up to a completely unstructured state $\delta_t > \pi ⁄2$ (see Materials and Methods).
Figure \ref{fig1}d reports the variation of $\delta_t$ within the cycle at $\gamma = 22.7\%$, clearly  showing the angle crossing the $\pi⁄4$ line, corresponding to partial yielding at this $\gamma$. A further increase in $\gamma$ in successive cycles results in $\delta_t$ plots where both the time during the cycle in which $\delta_t > \pi⁄4$, and its maximum value, increase.

\begin{figure}[t!]
\centering
\includegraphics[scale=1]{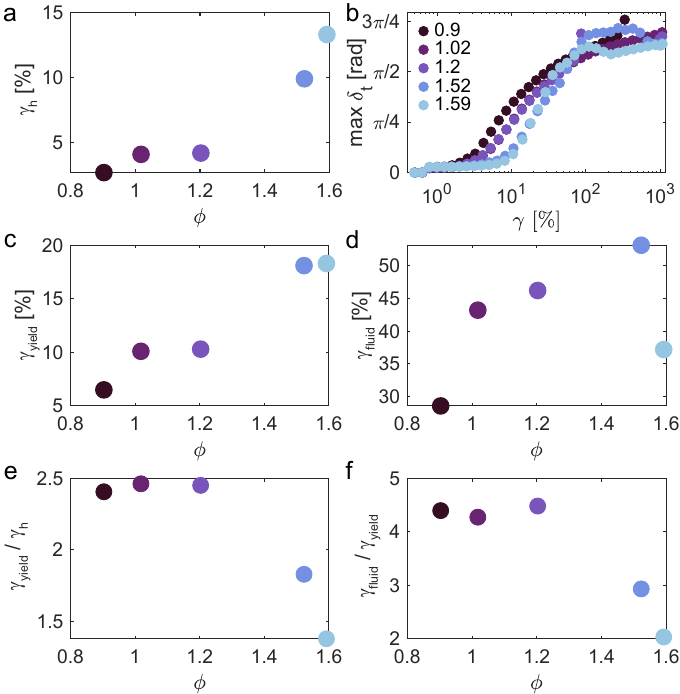}
\caption{\small \textbf{Time-resolved analysis of the rheological response of pN5 microgels.} a) $\gamma_h$ as a function of volume fraction. b) $max(\delta_t)$ as a function of strain amplitude in amplitude sweep experiments at increasing $\phi$. c) First strain value at which $\delta_t = \pi/4$, indicating partial yielding, as a function of volume fraction. d) First strain value at which $\delta_t = \pi/2$, corresponding to complete yielding, as a function of volume fraction. e) Strain ratio $\frac{\gamma_{yield}}{\gamma_h}$ as a function of volume fraction. f) Strain ratio $\frac{\gamma_{fluid}}{\gamma_{yield}}$ as a function of volume fraction. The full corresponding amplitude sweeps for each $\phi$ are reported in Figure S3.}
\label{fig2}
\end{figure}

To explore in details the yielding transition under oscillatory shear, we extracted three metrics from the time-resolved SPP analysis in order to identify when the material's response departs from a primarily solid-like behavior, then yields and finally fluidifies at very large applied deformations.
We first define $\gamma_h$ (Figure \ref{fig2}a) as the strain for which $max(G'_t) > 1.05 \, G'$ (see Figure S3). This metric allows us to unambiguously identify from which strain value the time-resolved analysis differs from the response of the harmonic approximation, indicating an appreciable sample reorganization within each deformation cycle.
The value of $max(\delta_t)$ (Figure \ref{fig2}b), i.e. the maximum of the $\delta_t$ vs strain curves obtained for each applied strain (see an exemplary curve in Figure \ref{fig1}d), is then used to summarize the time-resolved analysis and pinpoint at which strain values the material starts yielding, with $\gamma_{yield}$ corresponding to the first strain value at which $max(\delta_t) = \pi/4$ (Figure \ref{fig2}c). Finally, we identify the strain at which the material fully fluidifies ($\gamma_{fluid}$), corresponding to the first strain value at which $max(\delta_t) = \pi/2$ (Figure \ref{fig2}d).

The strain values related to these metrics increase as a function of $\phi$. This can be expected, as more concentrated samples become stiffer and can accommodate larger deformations before the internal structure breaks, and the sample yields and flows. A similar result is observed also when looking at the crossover point $G' = G''$ in the harmonic approximation (Figure S4a), typically considered as an indication of the yielding transition in the sample.\cite{Carrier2009,VanderVaart2013,Bonn2017,Ghosh2019,KaithakkalJathavedan2020} 
The values of $\gamma_{yield}$ (i.e., when $G'_t = G''_t$) are always lower than the corresponding $G' = G''$ ($\gamma_\mathrm{CR}$, Figure S4b): partial yielding within a deformation cycle happens before the actual crossover of the harmonic moduli, as already observed for similar samples. \cite{Donley2019} Interestingly, the distance between these two metrics, $\gamma_{yield}$ and $\gamma_\mathrm{CR}$, decreases as a function of $\phi$. We attribute the latter result to a more abrupt onset of the yielding transition upon increase in particle density.

We note that the apparent decrease of $\gamma_{fluid}$ at $\phi = 1.59$ is attributed to noise in the acquired data. 
When the same analysis was applied for strain amplitude sweeps at $\omega = 1 rad/s$ (Figure S5), the obtained value of $\gamma_{fluid}$ at $\phi = 1.59$ is comparable to that at $\phi = 1.52$ (Figure S5e). Figure S5 also shows that these results are independent of the oscillation frequency.

We then quantified the deformation span required for each transition to occur; in other words, the abruptness of yielding as a function of $\phi$. This is measured by the strain ratios $\frac{\gamma_{yield}}{\gamma_h}$ (Figure \ref{fig2}e) and $\frac{\gamma_{fluid}}{\gamma_{yield}}$ (Figure \ref{fig2}f), indicating, respectively, how much deformation of the sample is required to cause an appreciable modification of the internal structure until partial yielding occurs, and how much deformation can the material sustain before it completely yields.
Both ratios are approximately constant up to $\phi = 1.2$. In this volume fraction range, the transitions, even though all starting at higher absolute strain values, require a similar, relative, deformation range. This is visually captured also in Figure S3, where the $\gamma$ relative to each metric is superimposed on the $G'$, $G''$ moduli from LAOS.

A different behavior is instead observed at higher $\phi$, for which both ratios decrease. We attribute this effect to the extent of compression, deformation and interpenetration of the microgels in the structure as a function of $\phi$. In the extreme cases, $\phi > 1.2$ for this particular microgel, the particles are already significantly deformed at rest and can adapt less to the imposed sample deformation. 
As a consequence, the yielding transition, even though starting at higher (absolute) deformation values due to the increased sample stiffness, is sharper, i.e. takes place within a smaller deformation range. 
We point out that this behavior resembles that of hard spheres, for which the yielding transition occurs more abruptly than for softer particles. \cite{Koumakis2012}

\subsection{Effect of crosslinking density}

\begin{figure}[t!]
\centering
\includegraphics[scale=1]{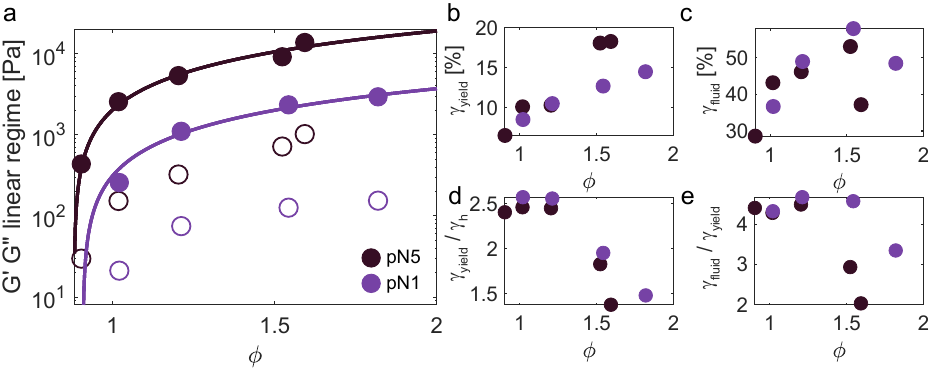}
\caption{\small \textbf{Effect of crosslinking density.} a) Storage (filled symbols) and viscous (open symbols) moduli in the linear regime ($\gamma = 0.5 - 1 \%$) as a function of $\phi$ ($\omega = 10$~rad/s). Solid lines represent linear fitting of the data. b) First strain value at which $\delta_t = \pi/4$ as a function of volume fraction. c) First strain value at which $\delta_t = \pi/2$ as a function of volume fraction. d) Strain ratio $\frac{\gamma_{yield}}{\gamma_h}$ as a function of volume fraction. e) Strain ratio $\frac{\gamma_{fluid}}{\gamma_{yield}}$ as a function of volume fraction.}
\label{fig3}
\end{figure}

We then performed the same analysis on pN1 microgels (Figures S6-S7) to quantify the influence of the single-particle internal elasticity on the rheological behavior in suspension. As for the stiffer pN5 microgels, the storage modulus increases linearly with $\phi$ at low deformations (Figure \ref{fig3}a). For comparable $\phi$, we observe, as expected, a decrease in $G'$, $G''$ values for microgels with lower crosslinker content, indicating softer samples at rest. \cite{Scotti2020}

The non-linear response as a function of $\phi$ is however more complex. In the lower $\phi$ range, we do not detect any appreciable difference in the onset of the yielding transition (Figure S8 and Figure \ref{fig3}b) and fluidization (Figure \ref{fig3}c) as a function of crosslinking density, despite the lower values of the corresponding linear moduli. Instead, for the more concentrated samples ($\phi > 1.5$) the increased softness of pN1 microgels shifts the onset of yielding to lower deformation values, but then complete fluidization requires larger strains (Figure \ref{fig3}c).  
This result is clearly captured by the strain ratios $\frac{\gamma_{yield}}{\gamma_h}$ (Figure \ref{fig3}d) and $\frac{\gamma_{fluid}}{\gamma_{yield}}$ (Figure \ref{fig3}e).
Higher values of $\frac{\gamma_{yield}}{\gamma_h}$ for pN1 microgels indicate that, at all the investigated $\phi$, softer particles can accommodate more (relative) deformation before yielding. This effect is even more pronounced when looking at high deformation values corresponding to complete fluidization, which show a significant broadening of the yielding transition.

We can conclude that samples made of softer microgels start to deform and rearrange at lower strain values as a consequence of the lower sample elasticity (lower $G'$, $G''$ values, associated with lower bulk moduli of the individual particles), however they can accommodate more (absolute) deformation before complete fluidization, presumably due to an increased ability of the particles to deform and interpenetrate to resist the applied deformation. Finally, for pN1 microgels, we also observe a decrease in the strain ratios (Figure \ref{fig3}d, \ref{fig3}e) at higher $\phi$. However, this decrease is shifted to higher volume fractions with respect to pN5. Softer microgels can reach higher effective volume fractions before the extent of single-particle deformation is so high that the microgels' ability to change their shape in response to the applied strain decreases and the deformation range between the onset of the yielding transition and complete fluidization progressively decreases.

\subsection{Influence of the polymeric shell}

\begin{figure}[t!]
\centering
\includegraphics[scale=1]{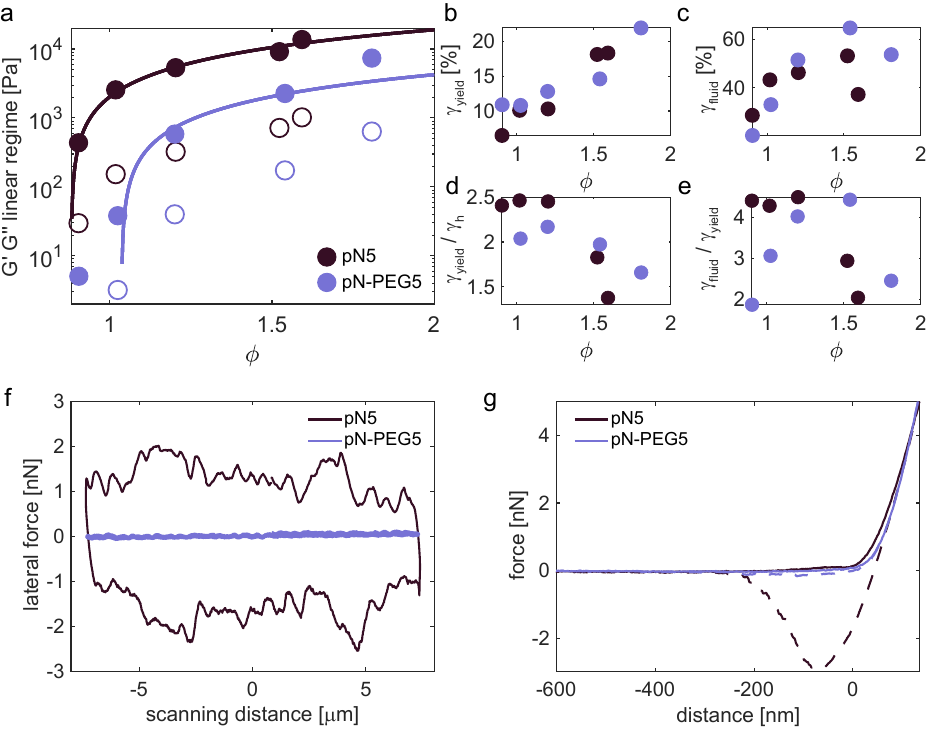}
\caption{\small \textbf{Influence of the polymeric shell.} a) Storage (filled symbols) and viscous (open symbols) moduli in the linear regime ($\gamma = 0.5 - 1 \%$) as a function of $\phi$. Solid lines represent linear fits to the data. b) First strain value at which $\delta_t = \pi/4$ as a function of volume fraction. c) First strain value at which $\delta_t = \pi/2$ as a function of volume fraction. d) Strain ratio $\frac{\gamma_{yield}}{\gamma_h}$ as a function of volume fraction. e) Strain ratio $\frac{\gamma_{fluid}}{\gamma_{yield}}$ as a function of volume fraction. f) Lateral force versus scanning distance (friction loops) for pN5 (black) and pN-PEG5 (violet) microgel monolayers deposited on a silicon wafer and immersed in water, obtained using colloidal probe atomic force microscopy (CP-AFM) at 25 °C under a normal load of 10 nN. See Materials and Methods for details. g) AFM force-vs-distance curves (approach curves solid lines, retraction curves dashed lines) obtained on the same microgel monolayers.}
\label{fig4}
\end{figure}

We then investigated the influence of surface forces on the rheological behavior of microgels in dense states by adding linear PEG chains on their surface, while keeping the crosslinker content unchanged.
We first note that the viscoelastic moduli of pN-PEG5 particles do not increase linearly with $\phi$ throughout the investigated regime (Figure \ref{fig4}a). A similar result was obtained on copolymer pNIPAM-PEG microgels, \cite{Lara-Pena2021} and attributed to a saturation of interpenetration and shell compression at the highest $\phi$, resulting in a more pronounced increase of $G'$, although the relatively low number of points here precludes a more detailed analysis of the rheological behavior at small deformations.
When comparing with pure pNIPAM microgels, the moduli are always lower in the presence of PEG for all $\phi$ investigated, albeit this difference decreases at the highest $\phi$.

Based on the DLS results and the comparable de-swelling curves of the two microgels, we hypothesize that changes in the rheological behavior are mostly dependent on interparticle interactions in crowded environments and not on single-particle softness, which remains similar for the same crosslinker content. In particular, PEG chains on the microgels' surface, being more hydrophilic than pNIPAM, might decrease adhesion and friction between particles, similarly to what observed in the case of pNIPMAM (poly(N-isopropyl-methacrylamide)) and pNIPAM colloids.\cite{Urayama2014} This effect is particularly relevant at lower $\phi$, when the microgels are less compressed and thus surface properties and interparticle contacts are expected to be more important in governing the rheological behavior, and it might explain the observed decrease of $G'$ and $G''$  for pN-PEG5 with respect to the homopolymer particles.

The influence of PEG-mediated surface interactions is also affecting the non-linear response of the samples under LAOS (Figure \ref{fig4}b-e, Figures S9-S10). Differences between pN5 and pN-PEG5 microgels stand out especially when looking at the ratios $\frac{\gamma_{yield}}{\gamma_h}$ (Figure \ref{fig4}d) and $\frac{\gamma_{fluid}}{\gamma_{yield}}$ (Figure \ref{fig4}e).
Up to $\phi < 1.2$, when the interactions between surfaces in contact dominate, the lower adhesion and friction between particles as mediated by the PEG chains results in samples that overall yield (Figure \ref{fig4}d) and finally flow (Figure \ref{fig4}e) in a lower deformation range. 
Instead, for $\phi > 1.2$, when the single-particle compression becomes more significant, the results resemble those obtained with pN1 microgels. However, while for pN1 microgels (Figure \ref{fig3}) the different yielding behavior was attributed to the lower crosslinker content, in this case, we ascribe the more extended yielding process to the fuzziness, and thus resulting higher compliance, of the particle shell. Higher shell compliance enhances the microgels' ability to accommodate imposed deformations by rearranging with their neighbors, and, as a consequence, the yielding process is shifted to higher deformations.

In order to rationalize these observations, we turned to colloidal probe atomic force microscopy (CP-AFM) to quantify the friction and adhesion of microgel monolayers with and without PEG. To this end, we used lateral force microscopy and nano-indentation on monolayers of pN5 and pN-PEG5 microgels deposited on a silicon support and immersed in water at 25°C. These measurements allow gaining information on the interfacial friction between the microgels' surface and the colloidal probe (silica colloid, diameter: 16 $\mu$m), as well as on the adhesion between the probe and the microgels. \cite{Liamas2020}
In Figure \ref{fig4}f we report friction loops (lateral force vs sliding distance) recorded by scanning the probe laterally over the microgel monolayers at a constant normal load of 10 nN. A comparable crosslinker content between the two samples rules out effects due to variations in the stiffness of the polymer layer, which in turn would affect the measured friction. \cite{Liamas2020} These data show that a very low friction is obtained in the case of monolayers of pN-PEG5, while pN5 microgels display a much higher friction, as visualized by the larger width of the friction loops. These observations are in agreement with previous studies indicating that PEG-coated surfaces tend to exhibit superior lubrication when compared to PNIPAM-coated surfaces in water at room temperature for the same grafting density, as measured by AFM. \cite{Brady2009,Ramakrishna2013,Morgese2018,Ramakrishna2017}

In a complementary set of experiments, we used the same samples and tip to perform adhesion tests by indenting the microgel monolayers and recording the resulting force versus distance curves. As reported in Figure \ref{fig4}g, we observe significant adhesion between the tip and the surface of pN5 microgels, while pN-PEG5 microgels display very low adhesion.
Overall, these results indicate that both adhesion and friction are significantly decreased for pN-PEG microgels, strongly support the assumption that PEG chains are mainly located on the particle surface, and corroborate the interpretation of the rheological behavior.

Similar results have been obtained for microgels with 1 mol \% crosslinker with and without PEG (Figure S11). Also in this case, the friction for microgels containing PEG chains is lower and the adhesion is significantly decreased, indicating that PEG chains are covering the particle surface in contact with the colloidal probe, modifying the microgels' surface properties.
We note that the increased friction between 5 mol \% and 1 mol \% samples can be attributed to the decreased stiffness of the microgel layer, and consequent higher deformation and resulting tip-sample contact area during compression with the colloidal probe. \cite{Liamas2020}
Regarding the rheological properties of pN1 and pN-PEG1 microgels (Figure S12), the results are overall coherent with what observed in the presence of a higher crosslinker content, although the response in the non-linear regime is too noisy for an accurate analysis due to the very low $G'$ values obtained for pN-PEG1.

\section{Conclusions}

In this work we investigated the rheological behavior of pNIPAM microgels under oscillatory shear in the linear and non-linear regimes, exploiting the sequence of physical processes (SPP) \cite{Rogers2017,Donley2019} approach to the analysis of LAOS data to gain insight on the materials viscoelastic response before, during and after the yielding transition upon increasing applied deformations. Overall, these results provide additional insight on the complex rheological properties of dense microgel suspensions, linking both particle elasticity (mediated by the crosslinker content) and interparticle contacts (dictated by the polymer chains on the particle surface) to the onset of yielding and to the overall transition from solid-like to fluid-like behavior. 

As already reported in previous works, \cite{Menut2012,Scotti2020} the linear viscoelastic moduli of dense microgel suspensions decrease as a function of particle softness, as a result of the reduced bulk modulus of individual particles. However, particle softness plays an opposite role in the non-linear response, as revealed by investigating the yielding process under LAOS. Yielding is found to be a continuous transition from a solid-like state at rest to a fluid-like state at large applied strains, spanning a broad range of strain amplitudes, with the onset and the complete fluidization of the sample that requires a larger amount of strain for softer microgels. This is attributed to the fact that softer particles can accommodate more deformation before the sample yields and ultimately fluidifies.

The incorporation of PEG chains on the particle surface can be used as an orthogonal way to lower the linear storage and loss moduli of the resulting dense suspensions. This is attributed to a decrease in the adhesion and friction between particles when the surface-to-surface interactions are mediated by polymer chains that are more hydrophilic with respect to pNIPAM, as quantified by using colloidal probe lateral force microscopy and nano-indentation experiments on microgel monolayers deposited on a solid support. When subjected to shear deformation, the lubrication provided by the PEG chains induces a faster yielding transition.

Finally, for all the investigated particles we observed that the samples at the highest packing fractions resist yielding up to higher deformations, but then fluidization is completed in a shorter deformation range. This indicates that, above a certain compression and deformation induced by crowding, the microgels decrease their ability to deform further when sheared, therefore resisting less to the imposed deformation. Such behavior is shifted to higher packing fractions both for loosely crosslinked particles and for pNIPAM-PEG microgels due to their increased softness and fuzziness, respectively.

Overall, we believe these results can be of great potential value in discerning contributions due to particle elasticity and interparticle contacts that govern the viscoelastic properties of dense suspensions of soft particles subjected to shear, providing tools to tune the mechanical and rheological behavior of such complex soft systems.

\FloatBarrier

\section{Materials and Methods}
\small
\subsection{Reagents}
N,N’-methylenebis(acrylamide) (BIS, Fluka 99.0\%), potassium persulfate (KPS, Sigma–Aldrich 99.0\%) and polyethylene glycol methyl ether methacrylate (PEGMA) m.w.: 13000 (abcr GmbH) were used without further purification. N-isopropylacrylamide (NIPAM, TCI 98.0\%) was purified by recrystallization in 40/60 v/v toluene/hexane. 

\subsection{Microgels synthesis}

\textit{pNIPAM microgels.} pNIPAM microgels were synthesized by surfactant-free semi-batch radical precipitation polymerization, following already published protocols.\cite{Vialetto2022} NIPAM (2g) and BIS were dissolved in 100 mL of MilliQ water. The amount of BIS was chosen in order to synthesize microgels with 5 or 1 mol \% crosslinker. The monomers mixture was purged with nitrogen for 1 h and then 60 mL of the monomer solution was taken out with a syringe. 20 mL of MilliQ water were added to the reaction flask and the solution was immersed into an oil bath at 80°C and purged with nitrogen for another 30 min. The reaction was started by adding 25 mg of KPS previously dissolved in 2 mL of MilliQ water and purged with nitrogen. Feeding of the monomer solution to the reaction flask was set at 0.5 mL/min and was initiated after 1.5 min. The reaction was quenched at the end of the feeding step quenched by opening the flask to let the air in, and placing it in an ice bath. The colloidal suspension was dialysed for a week, and purified by 6 centrifugation cycles and resuspension of the sedimented particles in pure water, and freeze-dried.

\textit{pNIPAM-PEG microgels.} pNIPAM-PEG microgels were synthesized by surfactant-free semi-batch radical precipitation polymerization, modifying already published protocols.\cite{Motlaq2018,Lara-Pena2021} NIPAM (1g) and BIS were dissolved in 100 mL of MilliQ water. The amount of BIS was chosen in order to synthesize microgels with 5 or 1 mol \% crosslinker, calculated with respect to all the monomers in the reaction mixture. Separately, PEGMA (1.2g) dissolved in 20 mL of MilliQ water and KPS (16 mg) dissolved in 2 mL of MilliQ water were purged with nitrogen for 1h. The reaction was carried out at 70°C and it was started by adding KPS to the reaction flask containing NIPAM. After 20 min, the PEGMA monomer was injected in the flask at 0.33 mL/min. The reaction was stopped after 5 h 20 min by opening the flask and placing it in an ice bath. The colloidal suspension was dialysed for a week, purified by 6 centrifugation cycles and resuspension of the sedimented particles in pure water, and freeze-dried.

\begin{table}[h]
\small
  \caption{\ Microgels used in this study, particle size at 19°C, swelling ratio and $k$} 
  \label{tbl:example1}
  \begin{tabular*}{1\textwidth}{@{\extracolsep{\fill}}llllll}
    \hline
    Microgel & mol \% BIS & mol \% PEGMA & d$_h$ (19°C) [nm] & d$_h$ (19°C) / d$_h$ (45°C) & $k$ \\
    \hline
    pN5     & 5 & - & 826 $\pm$ 10  & 1.77 $\pm$ 0.04 & 7.55 $\pm$ 0.2 \\     
    pN1     & 1 & - & 850 $\pm$ 8   & 2.16 $\pm$ 0.04 & 14.2 $\pm$ 0.2 \\     
    pN-PEG5 & 5 & 1 & 847 $\pm$ 20  & 1.78 $\pm$ 0.05 & 8.0  $\pm$ 0.1 \\    
    pN-PEG1 & 1 & 1 & 1013 $\pm$ 11 & 2.24 $\pm$ 0.05 & 21.6 $\pm$ 0.7 \\    
    \hline
  \end{tabular*}
  \label{tab1}
\end{table}

\subsection{Experimental methods}

\textit{DLS.} Dynamic light scattering (DLS) experiments were performed using a Zetasizer (Malvern, UK). The temperature was varied from 19 to 51°C with 2°C steps and 15 min equilibration time. For each temperature we recorded four consecutive measurements of 15 runs each. 

\textit{Viscometry.} Viscosity measurements were performed using a Micro-Ubbelohde viscometer (type no. 538 13, Xylem Analytics, Germany) immersed in a thermostat controlled water bath set at 25°C. We measured the relative viscosity of diluted microgel suspensions at different dilutions with respect to that of pure water. For each concentration, flow times were measured 5 times. The data were then fitted with the Einstein–Batchelor relation $\eta_{rel} = 1 + 2.5\phi + 5.9\phi^2$ with $\phi = k \cdot wt\%$, to determine the constant $k$ and the particle effective volume fraction $\phi$.

\textit{Colloidal probe AFM (CP-AFM).} 
Microgel monolayers were deposited onto silicon wafers via transferring from a hexane-water interface following procedures already described elsewhere. \cite{Vialetto2021,Vialetto2022JCIS} Briefly, microgels were first adsorbed at the hexane-water interface, where they formed ordered monolayers. A piece of a silicon wafer, previously cleaned with a UV-ozone cleaner (UV/Ozone Procleaner Plus, Bioforce Nanosciences) for 15 minutes to ensure a highly hydrophilic surface, was then lifted from the water subphase (where it was immersed prior to forming the microgel monolayer) through the interface at a constant speed of 25 µm·s$^{-1}$. This ensured transferring of the monolayer from the fluid interface to the solid substrate.
The silicon wafer was then lifted out of the hexane phase, left to dry and successively re-immersed in Milli-Q water for performing the AFM measurements.

Colloidal probe lateral force and adhesion force measurements were carried out using a MFP3D atomic force microscope (Asylum Research, Oxford Instruments, Santa Barbara, USA). The colloidal probe was prepared by gluing a silica colloid (16 $\mu$m in diameter, EKA Chemicals AB, Kromasil R, Sweden) to a tipless cantilever (CSC38, MikroMasch, Bulgaria) using a custom made micromanipulator. The prepared colloidal probe was treated under UV-ozone cleaner (Ossila, UK) for 20 minutes before the measurements. Friction loops were recorded by scanning the cantilever laterally over the microgel monolayer deposited on silicon, immersed in Milli-Q water, with a scan rate of 0.5 Hz and an applied load of $\sim$10 nN. The normal and torsional calibration of the cantilever was performed by using thermal noise \cite{Hutter1993} and Sader's method, \cite{Green2004} respectively. The lateral-force calibration was carried out by employing the ‘test-probe method'. \cite{Cannara2006} 

Adhesion measurements were performed using the same cantilever and on the same samples right after the friction tests. Force versus distance curves were measured at a scanning speed of 1 $\mu / s$.

\textit{Rheology.} Rheological measurements were performed using a Discovery HR-3 rheometer (TA Instruments), with a plate-plate geometry (stainless steel, diameter: 40 mm).  We note that, when using this configuration, no indications of responses associated with wall-slip is observed. 
A Peltier in contact with the lower plate ensured a constant temperature of 25°C (unless otherwise stated), and a solvent trap consisting of an enclosure with a solvent seal at the top and a wet tissue adhered to its interior was used to avoid evaporation. The gap in all experiments was set to 250 $\mu$m. A rejuvenation protocol was implemented before each measurement to minimize variabilities due to sample loading and aging. We first applied an oscillatory shear for 120 s with a large strain amplitude ($\gamma$ = 600\%) at frequency $\omega$ = 1 rad/s, during which all samples showed a liquid-like behavior. We then applied a second oscillatory shear with a low strain amplitude ($\gamma$ = 0.5\%) at frequency $\omega$ = 10 rad/s, until a steady state response in the viscoelastic moduli was reached. For all samples, such a steady state response was obtained within 120 s. Frequency sweep experiments were performed at $\gamma$ = 0.1\%, varying $\omega$ from 100 to 0.1 rad/s. Oscillatory shear experiments as amplitude sweeps were performed at constant frequency, with strain $\gamma$ varying from 0.5\% to 1000\%. For each point, we recorded 5 cycles; for each sample, we tested two frequencies, $\omega$ = 10 and 1 rad/s.

All time-resolved raw data were processed by using a freely available MATLAB-based software developed by Rogers.\cite{Donley2019} The data analysis is based on the sequence of physical processes (SPP) technique, which is used to define the instantaneous moduli ($G'_t$ and $G''_t$) throughout the time-varying response of the material to deformation, for each applied oscillatory shear. A complete derivation of the SPP parameters can be found elsewhere.\cite{Rogers2017,Donley2019} The SPP analysis was performed on data reconstructed via Fourier-domain filtering, using 5 harmonics. 
In short, for each oscillatory shear at fixed strain the analysis capture the instantaneous moduli $G'_t$ and $G''_t$, which describe the instantaneous response of the material within the cycle. The non-linear response of the material to deformation can be expressed in terms of the instantaneous magnitude of the viscoelastic response:
\begin{equation}\label{Gt*}
   |G*_t| = \sqrt{G'^2_t + G''^2_t}
\end{equation}
and the phase angle of the complex modulus:
\begin{equation}\label{delta_t}
   \delta_t = \tan^{-1} \left(\frac{G''_t}{G'_t}\right)
\end{equation}
As in the linear viscoelastic regime, yielding of a material can be identified on a macroscopic scale as the point within the cycle when the instantaneous response changes from primarily elastic, $\delta_t < \pi⁄4$, to primarily viscous, $\delta_t > \pi⁄4$. This correspond to a crossover between the time-resolved instantaneous moduli $G'_t$ and $G''_t$, analogous to the crossover between $G'$ and $G''$ in the linear approximation. As described in the works of Rogers\cite{Donley2019}, the range in which $\pi⁄4 < \delta_t < \pi⁄2$ can be considered as incomplete yielding of the material, when a significant degree of internal structure is maintained. Instead, complete yielding (completely unstructured state) can be defined as when the phase angle goes to the viscous limit of $\delta_t = \pi⁄2$. In this work we introduce another metric ($\gamma_h$) defined as $max(G'_t) > 1.05 \cdot G'$, to estimate when the rheological response within the deformation cycle starts to differ from the response in the linear approximation.

\normalsize


\section*{Conflict of Interest}
The authors declare no conflict of interest.

\section*{Data Availability Statement}
The data that support the ﬁndings of this study are available from the corresponding authors upon reasonable request.

\bibliography{ms_Arxiv}

\newpage

\begin{center}

\LARGE{Effect of particle stiffness and surface properties on the non-linear viscoelasticity of dense microgel suspensions\\
\bigskip
\Large{Supporting Information}
}
\end{center}

\section{Supporting Figures}

\begin{figure}[!htb]
\centering
\includegraphics[scale=1]{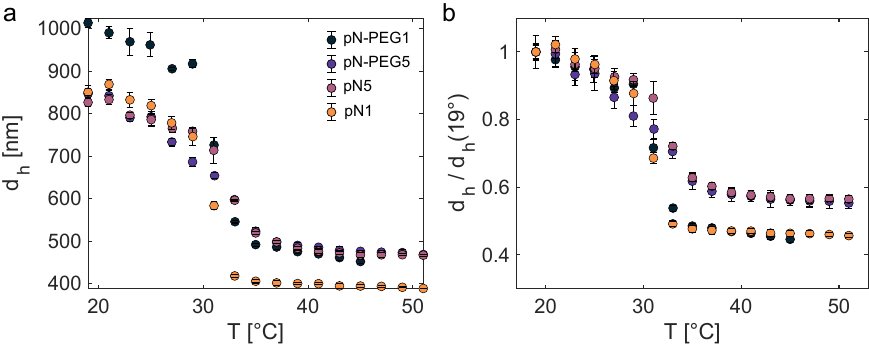}
\caption{\small \textbf{Microgels' hydrodynamic diameters as a function of temperature.} a) Hydrodynamic diameter $d_h$. Error bars indicate the standard deviation of 4 measurements. b) Swelling ratio as a function of temperature measured as $d = d_h/d_h(19^{\circ}C)$.}
\label{fig:DLS}
\end{figure}

\begin{figure}[!htb]
\centering
\includegraphics[scale=1]{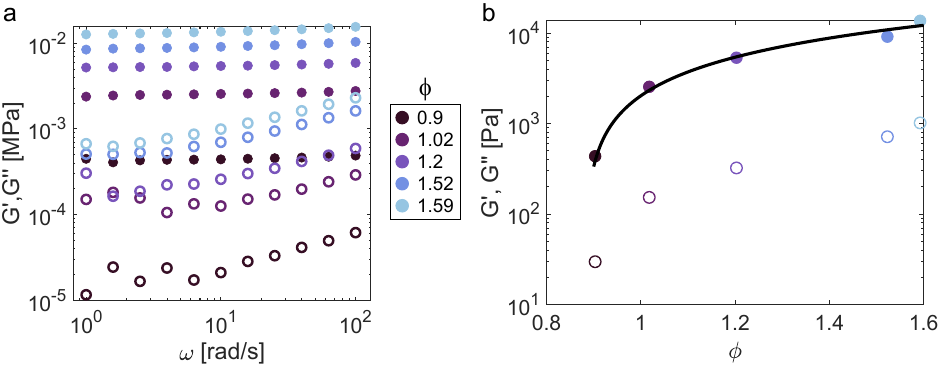}
\caption{\small \textbf{Linear rheology of pN5 microgels.} a) Dynamic frequency sweeps at $\gamma = 0.1\%$ for different values of $\phi$, as indicated. Solid symbols represent the storage moduli ($G’$) and open symbols the viscous moduli ($G''$). b) Plateau storage and viscous moduli as a function of $\phi$. The black line represent linear fitting of the data.}
\label{fig:ug45_linear}
\end{figure}

\begin{figure}[!htb]
\centering
\includegraphics[scale=1]{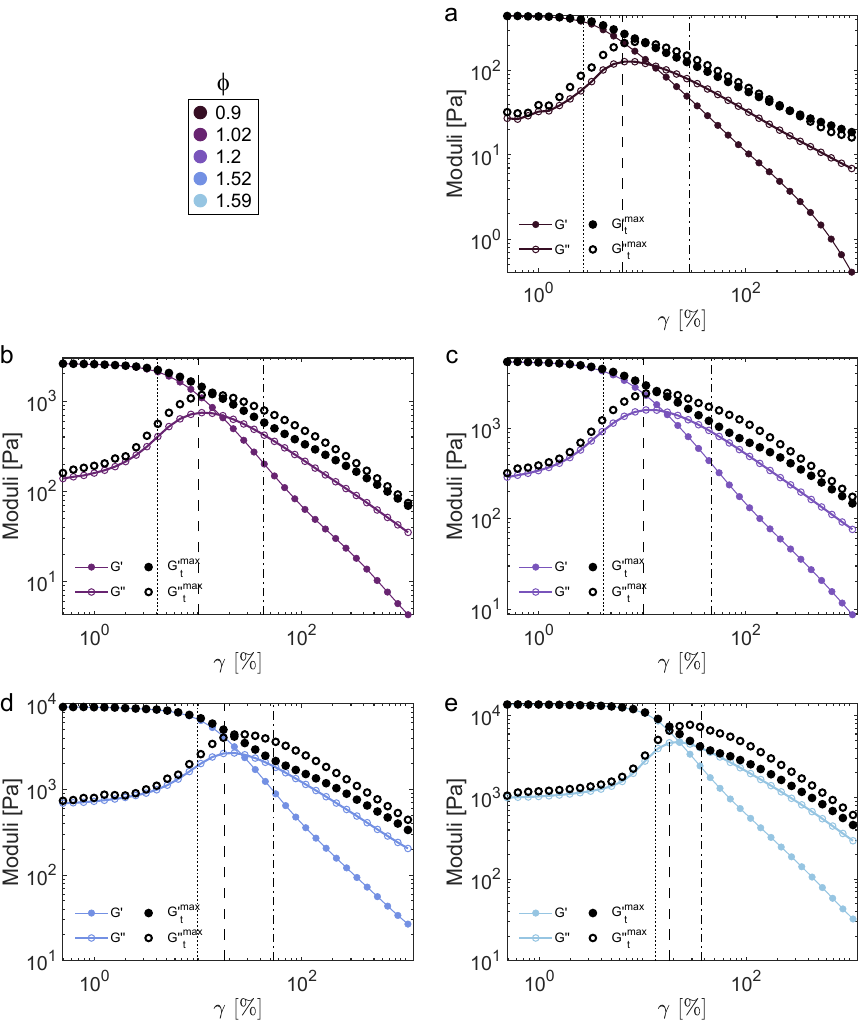}
\caption{\small \textbf{Large-amplitude oscillatory sweeps of pN5 microgels.} Amplitude sweeps as a function of strain amplitude $\gamma$, for an oscillation frequency of $\omega = 10 rad/s$, for different values of $\phi$. Color-coded symbols represent storage ($G’$, solid symbols) and viscous ($G’'$, open symbols) moduli; while black solid symbols $max(G'_t)$ and black open symbols $max(G''_t)$ for each strain amplitude. The dotted line indicates $\gamma_h$, the dashed line $\gamma_{yield}$, and the dash-dotted line $\gamma_{fluid}$.}
\label{fig:ug45_LAOS}
\end{figure}

\begin{figure}[!htb]
\centering
\includegraphics[scale=1]{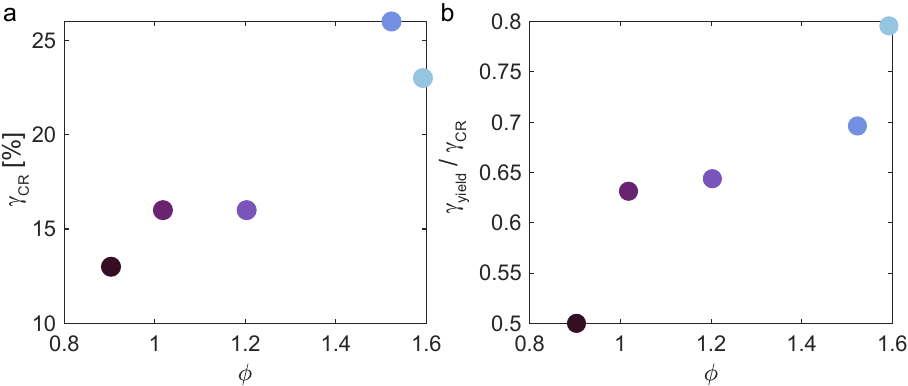}
\caption{\small \textbf{Crossover $G' = G''$ for pN5 microgels.} a) Strain value $\gamma_{CR}$ at the crossover $G' = G''$ point in the harmonic approximation, as function of $\phi$. b) Ratio $\frac{\gamma_{yield}}{\gamma_{CR}}$.}
\label{fig:ug45_crossover}
\end{figure}

\begin{figure}[!htb]
\centering
\includegraphics[scale=0.94]{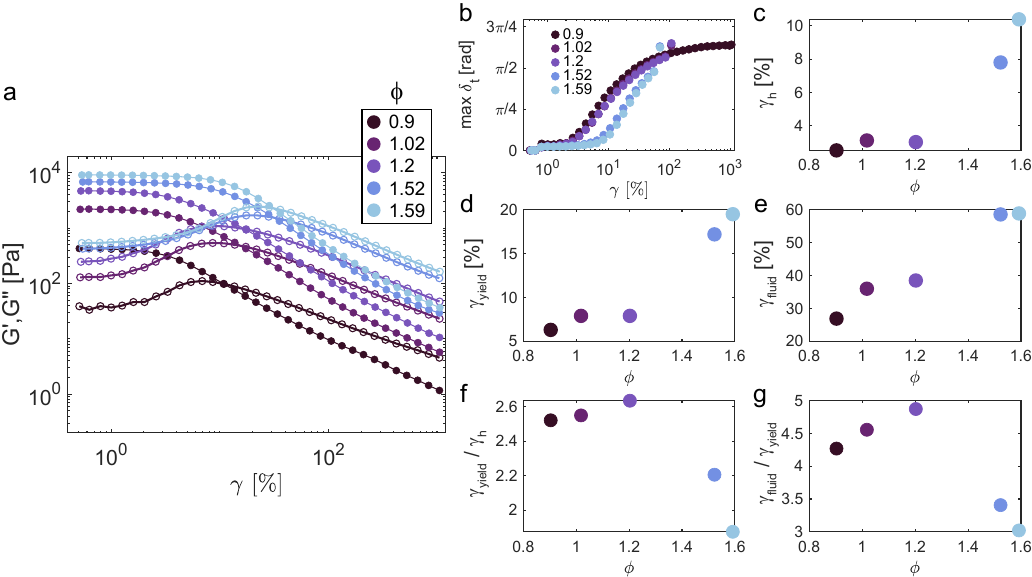}
\caption{\small \textbf{Large-amplitude oscillatory sweeps of pN5 microgels at $\omega = 1 rad/s$.} a) Amplitude sweeps as a function of strain amplitude $\gamma$, for an oscillation frequency of $\omega = 1 rad/s$, for different values of $\phi$. b) Plot of $max(\delta_t)$ in amplitude sweep experiments at increasing $\phi$. c) Strain corresponding to $\gamma_h$. d) First strain value at which $\delta_t = \pi/4$. e) First strain value at which $\delta_t = \pi/2$. f) Strain ratio $\frac{\gamma_{yield}}{\gamma_h}$. g) Strain ratio $\frac{\gamma_{fluid}}{\gamma_{yield}}$.}
\label{fig:ug45_1rads}
\end{figure}

\begin{figure}[!htb]
\centering
\includegraphics[scale=1]{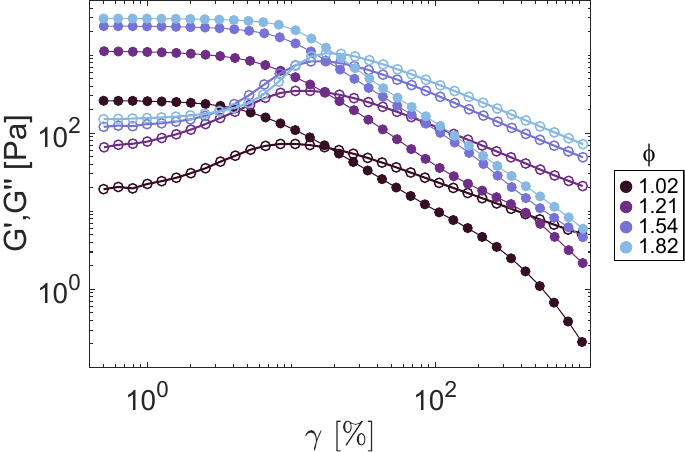}
\caption{\small \textbf{Large-amplitude oscillatory sweeps of pN1 microgels.} Amplitude sweeps as a function of strain amplitude $\gamma$, for an oscillation frequency of $\omega = 10 rad/s$, for different values of $\phi$.}
\label{fig:ug46_LAOS}
\end{figure}

\begin{figure}[!htb]
\centering
\includegraphics[scale=1]{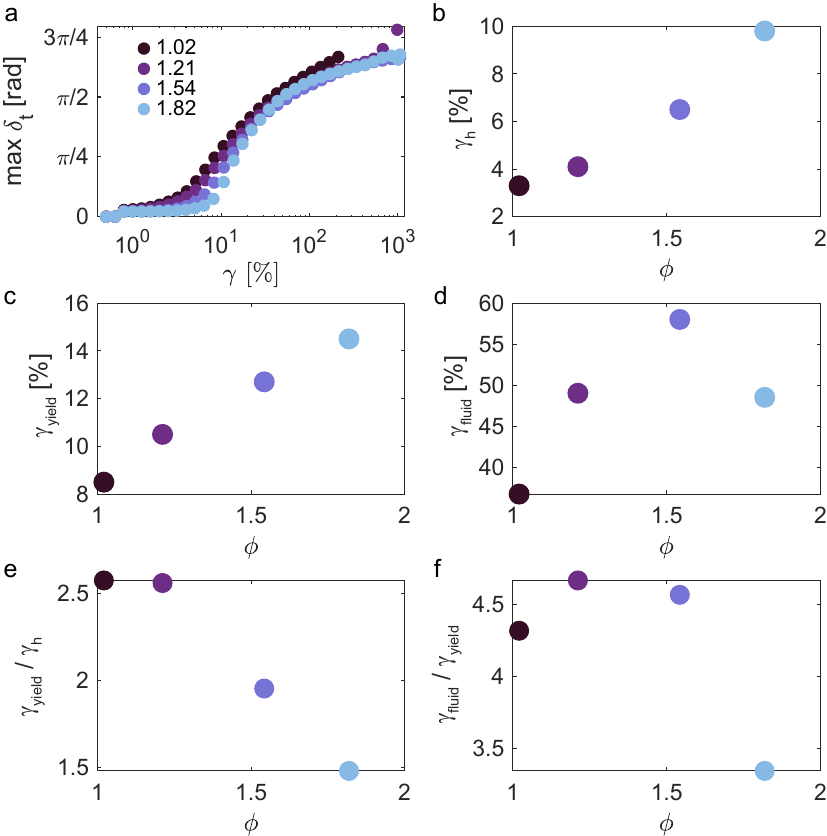}
\caption{\small \textbf{Time-resolved analysis of the rheological response of pN1 microgels.} a) Plot of $max(\delta_t)$ in amplitude sweep experiments at increasing $\phi$. b) Strain corresponding to $\gamma_h$. c) First strain value at which $\delta_t = \pi/4$. d) First strain value at which $\delta_t = \pi/2$. e) Strain ratio $\frac{\gamma_{yield}}{\gamma_h}$. f) Strain ratio $\frac{\gamma_{fluid}}{\gamma_{yield}}$.}
\label{fig:ug46_metrics}
\end{figure}

\begin{figure}[!htb]
\centering
\includegraphics[scale=1]{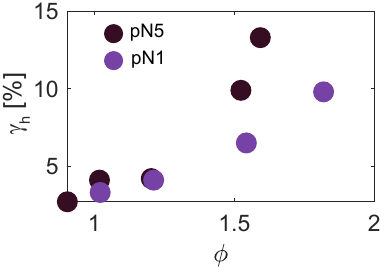}
\caption{\small \textbf{Strain $\gamma_h$ for pN5 and pN1 microgels.}}
\label{fig:ug45-ug46_oNL}
\end{figure}

\begin{figure}[!htb]
\centering
\includegraphics[scale=1]{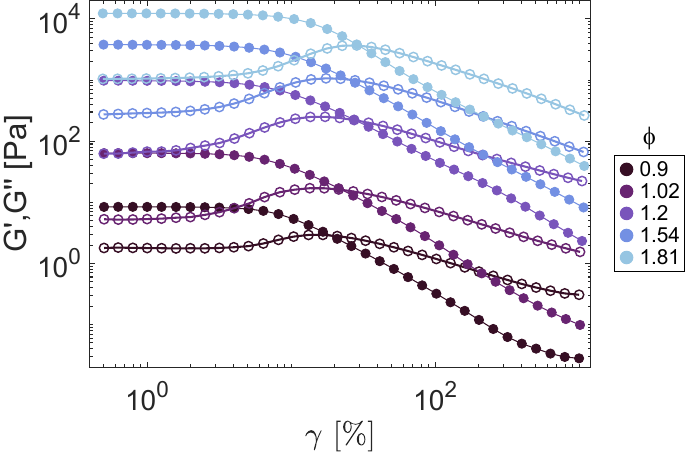}
\caption{\small \textbf{Large-amplitude oscillatory sweeps of pN-PEG5 microgels.} Amplitude sweeps as a function of strain amplitude $\gamma$, for an oscillation frequency of $\omega = 10 rad/s$, for different values of $\phi$.}
\label{fig:PEG13_LAOS}
\end{figure}

\begin{figure}[!htb]
\centering
\includegraphics[scale=1]{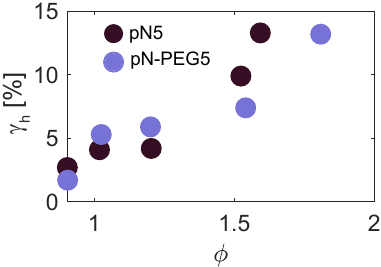}
\caption{\small \textbf{Strain $\gamma_h$ for pN5 and pN-PEG5 microgels.}}
\label{fig:ug45-PEG13_oNL}
\end{figure}

\begin{figure}[!htb]
\centering
\includegraphics[scale=1]{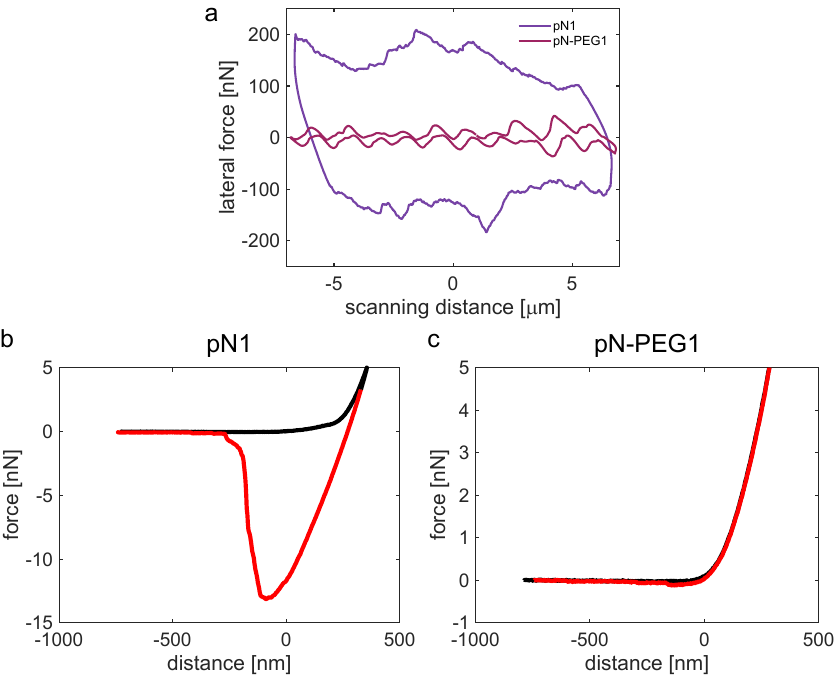}
\caption{\small \textbf{Friction loops and force vs distance curves obtained on 1 mol \% microgel monolayers.} a) Lateral force versus scanning distance for pN1 and pN-PEG1 microgels. See Materials and Methods for details. b-c) Representative AFM force-vs-distance curves (approach curve in black, retraction curve in red) obtained on a microgel monolayer re-swollen in water after deposition onto a silicon wafer. b) pN1 microgels. c) pN-PEG1 microgels.}
\label{fig:friction_1mol}
\end{figure}

\begin{figure}[t!]
\centering
\includegraphics[scale=1]{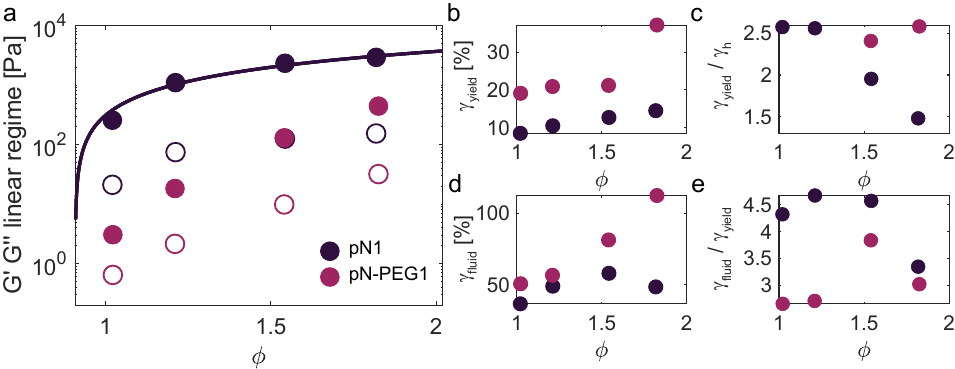}
\caption{\small \textbf{Comparison between pN1 and pN-PEG1 microgels.} a) Storage (filled symbols) and viscous (open symbols) moduli in the linear regime ($\gamma = 0.5 - 1 \%$) as a function of $\phi$. Lines represent linear fitting of the data. b) First strain value at which $\delta_t = \pi/4$. c) Strain ratio $\frac{\gamma_{yield}}{\gamma_h}$. d) First strain value at which $\delta_t = \pi/2$. e) Strain ratio $\frac{\gamma_{fluid}}{\gamma_{yield}}$.}
\label{fig:ug46_PEG08}
\end{figure}

\begin{figure}[!htb]
\centering
\includegraphics[scale=1]{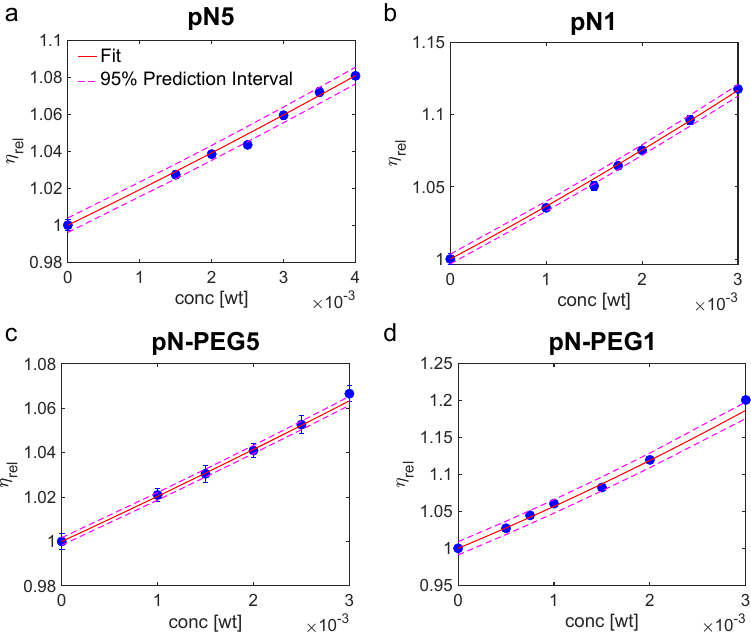}
\caption{\small \textbf{Experimental viscometry of dilute microgel suspensions.} Data are fitted with the Einstein–Batchelor equation (red line). Pink dotted lines represent 95\% prediction interval.}
\label{fig:viscosity}
\end{figure}



\end{document}